\documentclass[nopubldata]{lpor2014}
\usepackage{blindtext}
\usepackage{hyperref}
\hypersetup{%
  colorlinks,
  plainpages=false,
  pdfpagelabels,
  breaklinks,
  pdfview=Fit,
  bookmarksopen,
  bookmarksnumbered,
  linkcolor=black,
  anchorcolor=black,
  citecolor=black,
  filecolor=black,
  urlcolor=LPRred
  }%
\graphicspath{{./figs/}}
\setpages[]{}
\setvolume[0]{0}
\setyear{2011}%
\setdoi{201100000}%
\oddheadlogotext{}{Original paper}

\usepackage{setspace}

\usepackage{textcomp}

\usepackage{gensymb}

\newcommand{\heightimage}{3cm}

\newcommand{\widthminipage}{15cm}

\newcommand{\csub}{\mathrm{c}}
\newcommand{\FSR}{\mathrm{FSR}}

\newcommand{\isub}{\scriptsize \mbox{i}}

\newcommand{\msub}{\mathrm{m}}
\newcommand{\psub}{\scriptsize \mbox{p}}
\newcommand{\totsub}{\scriptsize \mbox{tot}}
\newcommand{\intens}{\mathfrak{I}}

\makeatletter
\newcommand\footnoteref[1]{\protected@xdef\@thefnmark{\ref{#1}}\@footnotemark}
\makeatother

\begin{document}

\titlefigure{Fig_Front_}

\abstract{
An external-cavity diode laser 
is reported with
ultra-low noise,
high power coupled to a fiber, 
and fast tunability.
These 
characteristics
enable the generation of 
an optical frequency comb 
in a silica micro-resonator
%
%
with a single-soliton state.
%
Neither an
optical modulator 
nor an amplifier
was used in the experiment.
%
This demonstration
greatly simplifies
the soliton generation setup 
and represents a significant step
forward to a fully integrated soliton comb system.

}


\title{
%
%
%
Micro-resonator soliton generated
directly with a 
diode laser
}

\titlerunning{
Micro-resonator soliton generated
directly with a 
diode laser
}

\author{
Nicolas Volet\inst{1}\textsuperscript{$\ast$},
Xu Yi\inst{2},
Qi-Fan Yang\inst{2},
Eric J. Stanton\inst{1},
Paul A. Morton\inst{3},
Ki Youl Yang\inst{2},
Kerry J. Vahala\inst{2},
and John E. Bowers\inst{1}
}
\authorrunning{N. Volet \textit{et al.}}
\mail{\email{volet@ece.ucsb.edu}}

\institute{%
Department of Electrical and Computer Engineering, 
University of California, Santa Barbara (UCSB), 
CA 93106, USA
\and
T. J. Watson Laboratory of Applied Physics, 
California Institute of Technology, 
Pasadena, CA 91125, USA
\and
Morton Photonics,
3301 Velvet Valley Drive,
West Friendship, MD 21794, USA
}

\keywords{
Single-mode lasers,
nonlinear optics, integrated optics,
temporal solitons.
}

\maketitle


\renewcommand{\heightimage}{50mm}
\newcommand{\widthimage}{50mm}
\renewcommand{\widthminipage}{8cm}


\begin{spacing}{.96} 

\section{Introduction}

Coherent optical frequency combs have 
revolutionized
precision  measurement 
with light 
\cite{Udem1999_1,Diddams2000,Jones2000}. 
However, 
the complex setups needed to generate
and stabilize
these systems 
limits their 
relevance
in field-testing scenarios.
%
Combs generated in a 
micro-cavity
\cite{DelHaye2007, Kippenberg2011} 
(``microcombs”)
have been evolving rapidly,
and
provide a pathway to miniaturize frequency comb
systems. While early microcombs were subject to
instabilities and lacked the critical ability to form
femtosecond pulses, an important advancement has been
the realization of temporal soliton mode locking in optical
microcavities 
\cite{Herr2014}.
%
First observed in optical fibers
\cite{Leo2010}, 
this form of mode locking has been demonstrated
across many microcavity platforms using
materials such as
silica (SiO$_2$) \cite{Yi2015}
and
silicon nitride (Si$_3$N$_4$) 
\cite{Brasch2016_1, Huang2016, Wang2016, Joshi2016}.
The potential to fully integrate frequency comb systems is
driving interest in semiconductor platforms that leverage
complementary metal-oxide-semiconductor (CMOS) 
fabrication infrastructures 
\cite{Heck2013, Roelkens2015}.
%
These are also
of interest
as soliton stabilization
\cite{Yi2016, Brasch2016_2}
necessitates 
certain electronic components.
%


%
%






Temporal 
optical
solitons 
\cite{Hasegawa1973_1}
were
first observed in 
optical fibers 
\cite{Mollenauer1980}.
%
These 
nonlinear
waves 
balance 
second-order
dispersion
using the 
optical Kerr effect.
For 
the new
temporal solitons in 
microcavities, an 
second balance 
also occurs
\cite{Herr2014}:
optical loss
is
compensated by 
third-order parametric gain
\cite{Kippenberg2004}
provided by 
an external pump.
The resulting dissipative Kerr soliton system is
therefore a regenerative mode-locked oscillator.
%
%
Soliton
mode locked microcavities have now been applied to
demonstrate dual-comb
spectroscopy \cite{Suh2016,YuM2017,Pavlov2017}, 
optical 
frequency synthesis
\cite{Spencer2017}, 
distance ranging 
\cite{Suh2017,Trocha2017} 
and optical communications 
with tremendous bandwidth
\cite{Marin-Palomo2017}. 
%


External-cavity diode lasers (ECDLs)
are 
critical
for many applications
and are frequently used to optically pump microcomb systems.
They can provide single mode and narrow linewidth,
which is crucial for telecommunications.
In addition, their emission frequency can be tuned,
which is needed 
for spectroscopy \cite{Schilt2004}
and frequency synthesis \cite{Spencer2017}.
However, they 
typically emit
modest output power.
%
Consequently,
%
all soliton 
microcomb
systems to date
require
non-integrated components 
to boost the pump laser power.
Additionally, other functions to enable
rapid control of laser power and frequency 
(acoustic/electro-optic modulators)  
\cite{Brasch2016_1,Yi2016,Stone2017arXiv}
have also been 
necessary.
In this work, 
we report an ECDL 
with extremely narrow linewidth,
low relative intensity noise (RIN),
high output power, and
useful spectral tunability.
As a very first application for this prototype, 
a temporal soliton 
is directly generated and stabilized 
in a 
high-$Q$ 
silica micro-resonator, 
%
%
without using an optical amplifier or an external modulator. 
This drastically reduces the complexity, size, 
and cost of the system,
and 
is expected to precede
a fully integrated source of solitons.



\end{spacing}


\section{Methods}
\label{SEC_METHOD}

\subsection{Summary of temporal soliton theory}
\label{SEC_2A}

%


Optical frequency combs 
can be generated by pumping 
a micro-resonator with 
a single-mode 
continuous-wave
(CW) laser
\cite{DelHaye2007}.
%
A
non-linear Schr\"odinger-like equation,
in its extended form 
\cite{Lugiato1987}
with driving, damping and detuning terms,
has proved remarkably successful 
in 
modeling
the dynamics of these
combs
\cite{Herr2014}.
%
%
Soliton solutions to 
this differential equation 
can be 
characterized by
the following temporal width 
\cite{Herr2014}:
\begin{align}
\label{Eq_tau}
\Delta\tau 
= \mathtt{C}_1 / \sqrt{ f_0 - f_{\psub} } ,
\end{align}
where 
$f_{\psub}$ is the laser frequency
and 
the resonator mode is centered at $f_0$.
%
%
The parameter
$\mathtt{C}_1 \equiv 
\sqrt{\frac{\vert \partial_{f} n_{\mathrm{g}} \vert}
{2 n_{\mathrm{g}}}} / (2 \pi)$
depends on the group index $n_{\mathrm{g}}$,
and its dispersion.
%
Note that 
a necessary condition for 
the generation of solitons 
is for the pump to be slightly red-detuned 
relative to the resonance 
($f_{\psub} < f_0$)
\cite{Coen2013-2}.
However,
due to 
the thermo-optic effect,
stable comb generation 
can be
obtained only by 
approaching the pump frequency
from the blue side of the resonance
\cite{Carmon2004}.
%
%
%
%
%
%
%
%
%
The upper envelope of the soliton 
spectral power density
can be approximated by
\cite{Yi2016_2}:
\begin{align}
\label{F_1}
\intens\left( f \right)
\propto 1 /
\cosh^{2}  \left[ \pi^2  
\left( f - f_{\psub} + \delta \right) 
\Delta\tau
\right] ,
\end{align}
where
$\delta$ is the soliton self-frequency shift
\cite{Yi2015}.
%
The spectral
maximum occurs at $f = f_{\psub} - \delta$.
Indeed, the maximum of the soliton power 
does not necessarily coincide with the pump frequency.
Integrating Eq.~(\ref{F_1})
and using Eq.~(\ref{Eq_tau}),
the time-averaged soliton power 
follows
\cite{Yi2015}:
\begin{align}
\intens_{\mathrm{sol}} 
\propto
\sqrt{f_0 - f_{\psub} } .
\label{I_sol}
\end{align}
%
%
This one-to-one relation between soliton power and the
pump-resonance detuning has been used to servo lock the
pump laser to the cavity by
fixing the soliton power
\cite{Yi2016}.
%
%
From Eq.~(\ref{Eq_tau}),
this 
procedure
also 
sets
the pulse temporal width.

\subsection{Ultra-low-noise, high-power ECDL}
\label{SEC_HL}

\begin{figure}[h!]
\begin{center}
\raisebox{17mm}{(a)}
\includegraphics[width=75mm]{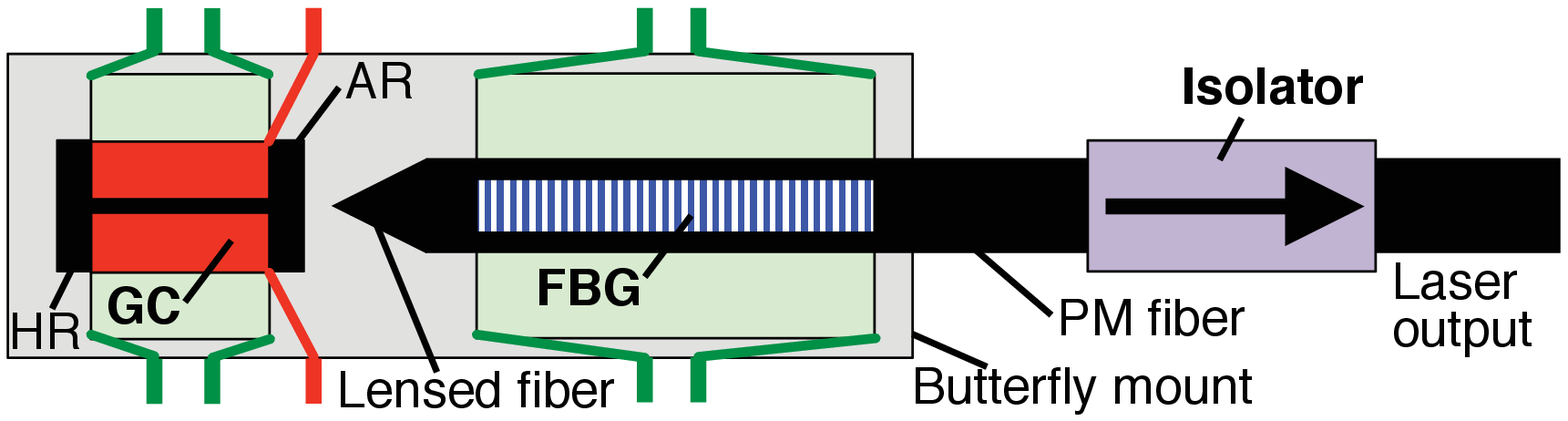}
\\
\vspace{2mm}
\raisebox{43mm}{(b)}
\includegraphics[height=46mm]{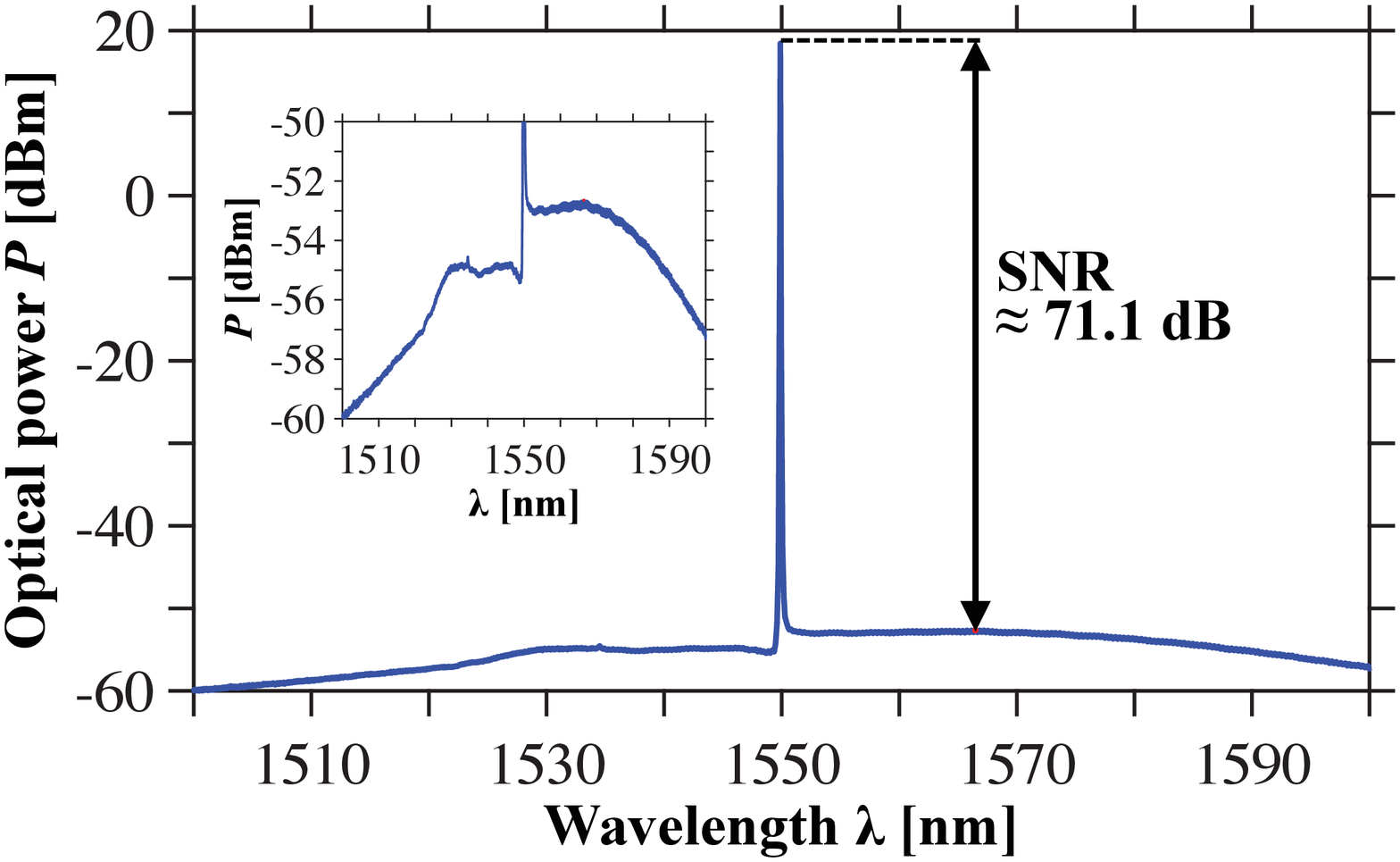}
\\
\vspace{2mm}
\raisebox{47mm}{(c)}
\includegraphics[height=50mm]{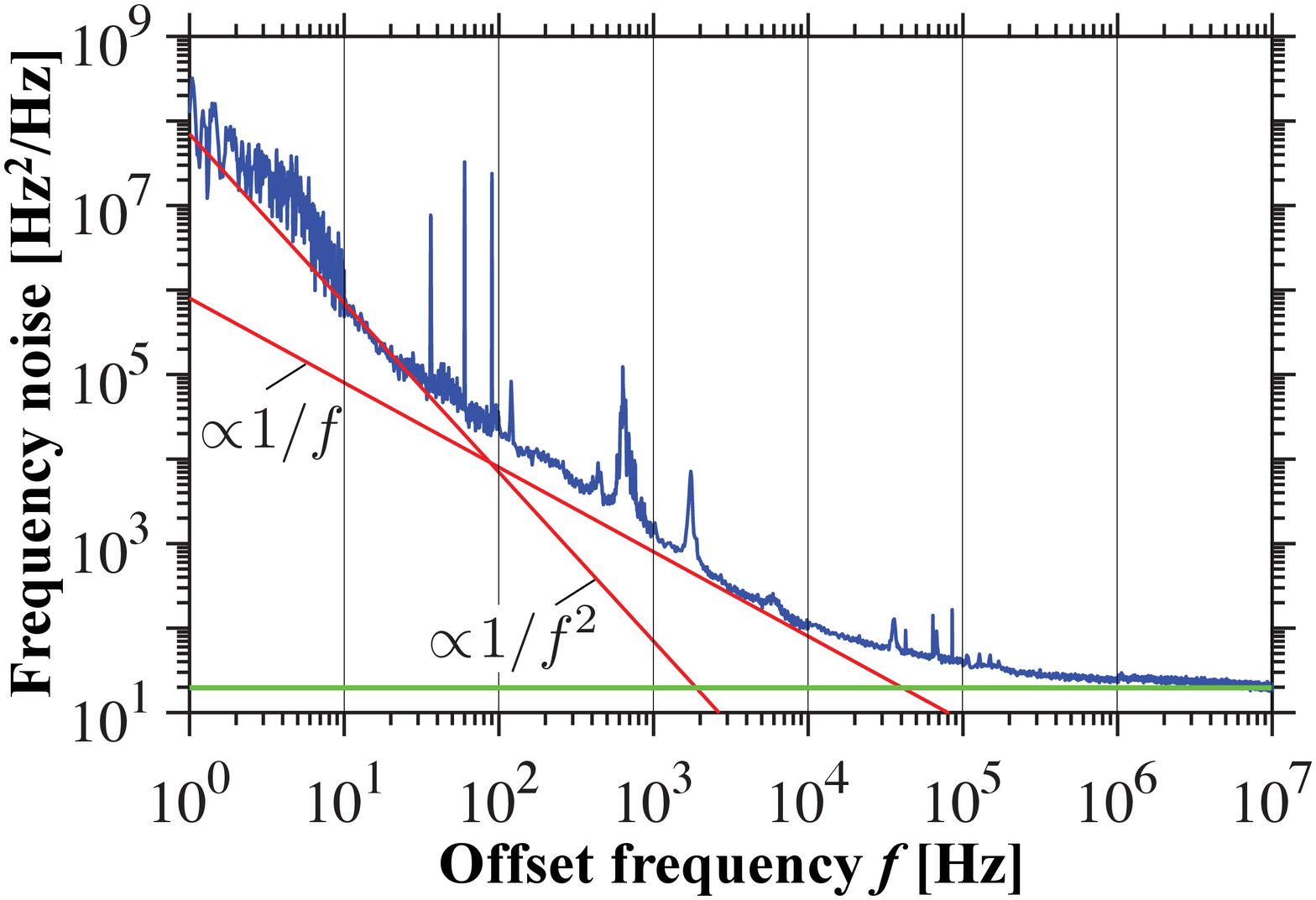}
\caption{
(a) Top-view schematic of the ECDL used in this work.
Pads of the butterfly package drawn in red 
are for current injection to the GC,
and those in green are for temperature control.
%
(b) Optical spectrum measured at 525 mA and 
with $T_{\mathrm{GC}} = T_{\mathrm{FBG}} =$ 20\,$^{\circ}$C.
The inset
is close-up view of the 
amplified spontaneous emission.
(c) Corresponding frequency noise.
Red lines are proportional to $1/f$ and $1/f^2$,
respectively,
and the green line 
indicates the white noise floor.
}
\label{FIG_M_02}
\end{center}
\end{figure}

A schematic of the 
prototype
ultra-low noise (ULN) 
ECDL
\cite{Morton_website_short}
is shown in Fig.~\ref{FIG_M_02}(a).
It combines
a 
high performance
semiconductor 
gain chip (GC) and 
a polarization-maintaining (PM)
optical fiber 
with an integrated 
custom designed
fiber Bragg grating 
(FBG)
\cite{Bird1991},
which forms one end of
the ECDL.
The GC 
consists of 
%
a multi-quantum well (MQW) active region
grown on an InP substrate.
%
Dielectric layers 
are deposited 
on one 
facet
to form a high-reflectivity (HR) 
coating, defining
the other end 
of 
the ECDL
\cite{Morton1994}. 
%
The opposite end of the GC has both an angled
waveguide and an anti-reflection (AR) coated facet 
to provide extremely low optical reflectivity,
suppressing any parasitic Fabry-Perot reflections 
and supporting ULN operation.
The light emitted from the GC
is coupled 
to the PM fiber via 
an AR coated 
lensed fiber.
%


\renewcommand{\widthimage}{55mm}
\renewcommand{\heightimage}{45mm}


The 
ECDL
design is optimized to provide 
ULN performance together 
with extremely stable single-mode operation 
\cite{Morton2017}.  
The temperature 
of the GC and of the FBG
are independently controlled,
which allows for coarse frequency tuning.
They are 
packaged
in an extended butterfly 
mount.
%
The present ECDL
does not include any moving parts
and its footprint is greatly reduced
compared to 
benchtop 
ECDLs
with a piezo-controlled frequency tuning mechanism
\cite{Cook2012}.
%
These are inherently less robust,
as their 
cavity alignment
is very sensitive to 
temperature fluctuations and vibrations.
%
%
%
A PM
optical isolator 
(OZ Optics) with $\sim$60-dB isolation
is spliced 
to the output fiber
to preserve the stability of this device
against 
optical reflections.
Note that it has recently been suggested 
that 
back reflections 
from the micro-resonator 
could be leveraged
\cite{Taheri2017},
so that 
an isolator may not be necessary 
for soliton generation.
This ECDL has a threshold near 40 mA,
and an output power of 70 mW 
was measured 
after the isolator
with a drive current of 525 mA.
%
%
At this current level, the 
signal-to-noise ratio (SNR)
is $>$70 dB, as seen 
in Fig.~\ref{FIG_M_02}(b).
%
The inset of Fig.~\ref{FIG_M_02}(b)
is a close-up view 
of
the amplified spontaneous emission (ASE).
On the red side of the 
lasing peak,
the ASE is $\sim$1.5 dB higher than on the blue side.
This asymmetry 
can be 
understood from 
the non-linear interaction 
between 
the optical field
and the carrier density in the QWs
\cite{Bogatov1975,Kalagara2013}.
Specifically, 
%
the beating 
between the intense lasing mode
and the ASE
modulates the (complex) refractive index,
which induces extra gain at longer wavelengths.


The frequency noise was measured
at 525 mA 
with an automated system 
(OEwaves, model OE4000).
As seen in Fig.~\ref{FIG_M_02}(c),
a white noise floor of 
$\sim$20 Hz$^2$/Hz
is obtained for frequencies
greater than $\sim$10$^6$ Hz.
This relates \cite{Kikuchi1985} to 
a Lorentzian 3-dB linewidth of 
$\sim$63 Hz.
The 
low 
linewidth
of 
this laser is
due to its long external cavity length,
high storage of photons 
within this cavity \cite{Henry1983}, 
and operation on the long-wavelength side 
of the FBG reflection 
\cite{Kazarinov1987, Vahala1984}.  
%
%
In the frequency range 30 Hz-2 kHz,
where the $1/f^2$-noise
and the $1/f$-noise dominate, 
several distinct peaks
can be seen in Fig.~\ref{FIG_M_02}(c).
They are inherent to the noise from
the electronics of the OEwaves measurement system.
%
On the other hand, contributions
in the range 1-10 Hz 
come
from the temperature controllers.


%
%
%
%
RIN measurements were also 
performed using the OEwaves OE4000 system.
Over the entire 
offset frequency
range from 10 Hz to 100 MHz, 
the measured RIN is equal 
to the noise floor of the equipment, 
thus only providing an upper limit 
to the RIN of the ECDL.
The noise floor of the OE4000 unit varies 
from $-130$ dBc/Hz at 10 Hz, 
down to almost $-160$ dBc/Hz at 100 kHz, 
and is then 
flat out to 100 MHz.
This RIN performance is comparable to
state-of-the-art low-noise ECDLs
in the C band
\cite{Loh2011},
while the linewidth demonstrated 
by the device in 
the present
experiments
is 
$\sim$15 times narrower.

%




%


\renewcommand{\widthimage}{55mm}
\renewcommand{\heightimage}{45mm}
\begin{figure}[h!]
\begin{center}
\raisebox{47mm}{(a)}
\includegraphics[height=50mm]{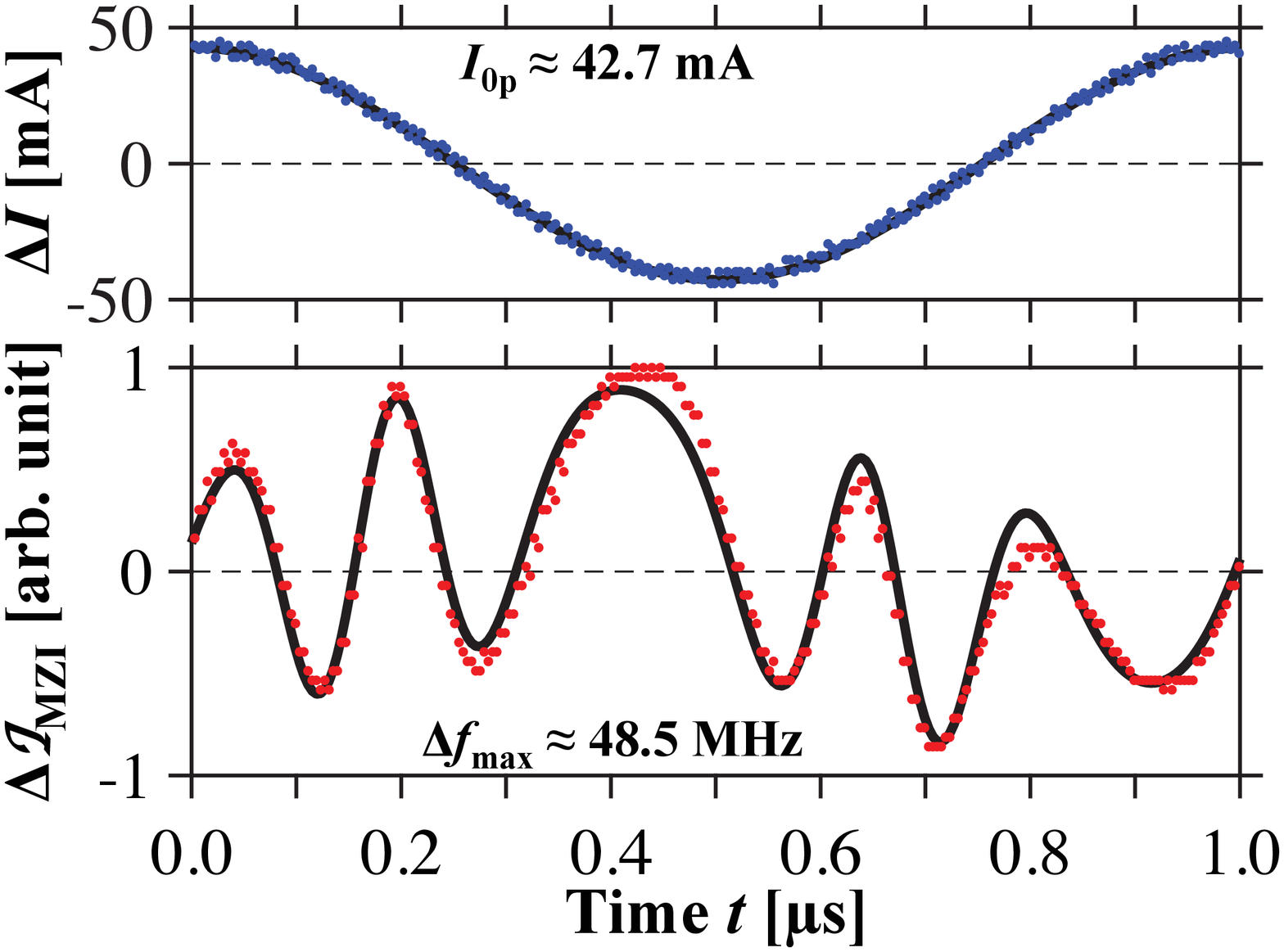}
\\
\vspace{2mm}
\raisebox{47mm}{(b)}
\includegraphics[height=50mm]{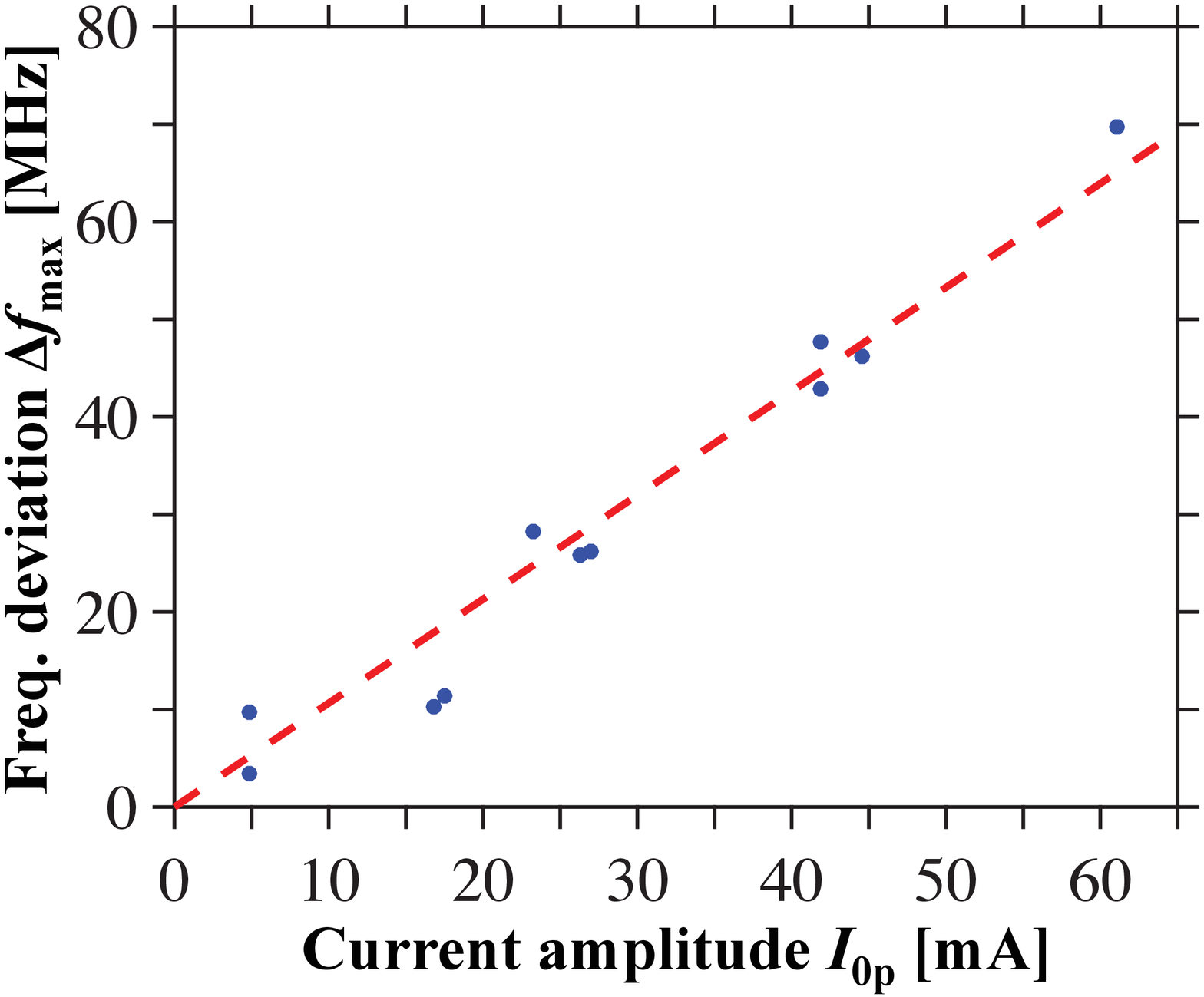}
\caption{
(a) 
Upper,
small-signal 
current modulation
of the ECDL 
for a certain value of 
$I_{\mathrm{0p}}$,
and 
lower,
intensity out of the MZI
fitted (in black) with Eq.~(\ref{EQ_4_all}).
(b) 
Peak
frequency deviation 
$\Delta f_{\max}$
obtained 
for different current amplitudes 
$I_{\mathrm{0p}}$.
The FM efficiency
is extracted from the slope of 
the linear fit (in red).
}
\label{FIG_M_FM}
\end{center}
\end{figure}


To characterize the tuning range, 
the ECDL output 
was
sent 
through 
a fiber-based
unbalanced Mach-Zehnder interferometer 
(MZI) 
with a free spectral range (FSR) of $\sim$40 MHz.
By 
slowly
ramping the current in the ECDL, a normalized
frequency change of $\sim$20 MHz/mA is measured,
and the widest
mode-hop-free tuning range is 
2.28 GHz (or 18.3 pm).
This is much narrower than 
specified for benchtop ECDLs, 
but as will be shown below, 
the tuning speed is at least 2 orders higher.
%
%
%
To characterize the possible 
tuning speed,
a 
time-harmonic
voltage
was
applied 
to 
the ECDL,
with a modulation frequency
$f_{\msub} = \omega_{\msub} / (2 \pi)$.
The current
through the ECDL
can be expressed as
$I(t) = I_0 + \Delta I(t)$,
with 
a mean value $I_0$,
a modulation
$\Delta I(t)
= I_{\mathrm{0p}} \cos \left( \omega_{\msub} t \right)$,
and a current peak amplitude
$I_{\mathrm{0p}}$.
%
%
This 
causes
the laser frequency 
to vary harmonically,
with a peak deviation $\Delta f_{\max}$.
For 
$f_{\msub} \ll \Delta f_{\max}$
and 
small-signal modulation 
($I_{\mathrm{0p}} \ll I_{0}$),
the 
intensity out
of the MZI can be written as
\cite{Schilt2004}:
\begin{subequations}
\label{EQ_4_all}
\begin{align}
\intens_{\mathrm{MZI}}
=
\intens_{0}
+ \Delta \intens_{\mathrm{MZI}}
= \left( \intens_{0}
+ \Delta \intens_{\mathrm{IM}} \right)
\left(
1 + \Delta \intens_{\mathrm{FM}} / \intens_{0}
\right) ,
\end{align}
where
$\intens_{0}$ is the unmodulated intensity,
and:
\begin{align}
\Delta \intens_{\mathrm{IM}}
&= m \intens_{0} 
\cos \left( \omega_{\msub} t 
+ \varphi_{\mathrm{IM}} \right)
\label{EQ_4b}
\\
\Delta \intens_{\mathrm{FM}}
&= R \intens_{0} 
\cos \left[ 2 \pi \frac{\Delta f_{\max}}{\FSR} 
\cos \left( \omega_{\msub} t + \varphi_f \right) 
+ \varphi_{\mathrm{FM}} \right] .
\label{EQ_4c}
\end{align}
\end{subequations}
Eq.~(\ref{EQ_4b}) represents
the spurious intensity modulation (IM)
as the laser frequency is chirped,
and $m$ is the IM index.
In contrast,
Eq.~(\ref{EQ_4c}) 
is used to extract the FM efficiency:
$\eta_{\mathrm{FM}} 
\equiv \Delta f_{\max} / I_{\mathrm{0p}}$.
The parameter 
$R$ 
is related to the MZI couplers,
and the $\varphi$'s are phase 
constants.
%
%
In the measurement,
the amplitude $I_{\mathrm{0p}}$ is varied up to 61 mA
and the modulation frequency is set to 
$f_{\msub} = 1$ MHz.
Note that for $f_{\msub}$ below 50 MHz,
temperature-induced FM effects 
are expected to be significant 
\cite{Kobayashi1982,Bowers1985}.
%
Specifically,
$\eta_{\mathrm{FM}}$
is expected to decrease as 
$f_{\msub}$ increases from DC to 50 MHz.
It is then usually relatively constant
up to the relaxation resonance.
An example is shown in 
Fig.~\ref{FIG_M_FM}(a) 
for $I_{\mathrm{0p}} = 42.7$ mA.
Fig.~\ref{FIG_M_FM}(b) 
demonstrates frequency tuning speed 
above
280 MHz/\textmu s.
%
%
Significantly,
a frequency deviation of 40 MHz
(required for soliton generation in a silica micro-disk),
can be generated with
a current amplitude of 40 mA.
Data 
in
Fig.~\ref{FIG_M_FM}(b)
are well fitted with a straight line,
and the slope is 
%
$\eta_{\mathrm{FM}} \approx 1.1$ MHz/mA.



\subsection{Silica micro-resonator}

The thermally-grown silica 
micro-disk resonator
is fabricated on a Si substrate 
with a technique reported 
elsewhere \cite{Lee2012}.
Its diameter is 3 mm, which corresponds
to an FSR of 
22 GHz.
%
The resonator 
profile
is a wedge
with an angle near 30$^{\circ}$
and a thickness of 8 \textmu m.
This geometry can simultaneously 
provide anomalous GVD 
\cite{Li2012a,Yi2015},
high $Q$ and minimal mode crossing.
The latter effect
is known to 
hinder soliton formation
\cite{Herr2014_2}.
Light is evanescently coupled 
to
the micro-resonator via a tapered fiber
\cite{Cai2000,Spillane2003}.
%
%
From transmission measurements,
the resonance used 
in this work
for soliton generation
has a 
loaded,
full-width-at-half-depth 
(FWHD)
linewidth
of 
1.2 MHz,
and therefore 
a total quality factor
$Q_{\totsub} \approx 160$ M.
The micro-resonator was under-coupled
in the measurement and had
an intrinsic value 
$Q_{\isub} \approx 260$ M.


To overcome the thermo-optic effect so that the pump
laser frequency is red-detuned relative to the cavity
resonance for soliton generation requires rapid tuning over
frequency spans as large as 40 MHz at rates in the range of
0.1-2 MHz/\textmu s \cite{Yi2016}. 
This range and rate are readily
achievable with the ECDL. 
As an aside, rates in the range
of 0.3-3 MHz/\textmu s are required 
in SiN combs \cite{Brasch2016_2} and
also of reach for the ECDL.


\subsection{Setup for soliton generation}

\begin{figure}[h!]
\begin{center}
\begin{minipage}[b]{\widthminipage}
\centering
\includegraphics[width=80mm]
{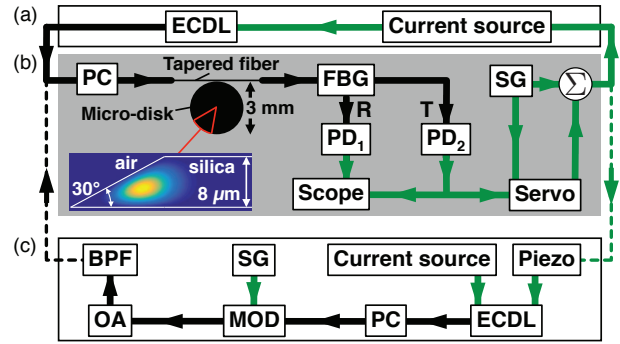}
\caption{
(a)-(b)
Schematic of the setup 
used in this work
for soliton generation and 
locking
without an OA.
Black and green lines 
represent optical and 
electrical
paths, 
respectively.
%
%
A cross-section schematic of the micro-disk
resonator and its wedge profile is shown in (b), 
with intensity pattern
simulated for the fundamental mode.
Alternatively,
for soliton generation with 
a piezo-controlled ECDL,
part (a)
is to be replaced 
by part (c).
}
\label{FIG_setup}
\end{minipage}
\end{center}
\end{figure}

Fig.~\ref{FIG_setup}(a)-(b)
shows a schematic of the setup used 
for soliton generation and 
locking
with 
the ECDL
described 
in Section~\ref{SEC_HL}.
%
%
The ECDL is driven 
by a low-noise current source 
(Newport
LDX-3620B).
%
A signal generator 
(SG, Keysight 33522B)
produces a 
local oscillation (LO)
used to modulate
the voltage supplied to the ECDL,
and thereby its emission frequency.
The output of the ECDL is connected to
a polarization
controller (PC, Thorlabs FPC560),
%
%
%
%
%
%
%
%
%
%
and 
the output of the tapered fiber
is connected to
a 
FBG (AOS).
%
%
Its reflection port (R),
with a 
0.1-nm 
pass-band,
transmits the pump laser 
light
which is sent to a photo-detector 
(New Focus 1811, labeled PD$_1$)
%
and
monitored
on an oscilloscope.
%
In contrast, 
the transmission port (T) of the FBG
acts as a notch filter and 
suppresses 
the pump 
by 25 dB,
so that
only the light produced by 
comb emission
and the ASE
are detected by
PD$_2$.
Part of this photo-current is
recorded on the oscilloscope,
and the 
remainder 
is sent to
a servo controller 
(Vescent D2-125)
that 
subtracts
a 
constant
offset.
The resulting difference 
serves
as the error signal
for a feedback-locking loop
\cite{Yi2015,Yi2016}
essentially based on Eq.~(\ref{I_sol}).

Once the pump frequency reaches 
the red side of the resonance
(prerequisite for soliton generation, 
see 
Section~\ref{SEC_2A}),
the SG sends a digital signal 
to 
the servo 
to engage 
the locking loop.
%
%
The servo controller 
includes 
an op-amp 
integrator.
To establish a stable loop,
it
forces a null 
at its input by integrating the error signal 
within a 
10-kHz 
bandwidth.
Its output is 
a correction voltage 
signal
that is fed back to 
the ECDL current source,
combined with the LO.
%
%
%
%
%
%
%
By 
adjusting 
the offset value,
states with different soliton numbers 
can be generated
\cite{Yi2016}.
As shown in the next Section,
the tuning and locking parameters 
were 
optimized to guarantee 
the generation 
and stabilization
of a \emph{single}-soliton state.


The conventional methods for generation of solitons using
piezo-controlled ECDLs can be understood 
by replacing 
part (a) 
with part (c) in Fig.~\ref{FIG_setup}. 
Specifically, 
two common methods 
have been
developed to overcome the thermal
instability
and generate solitons.
%
They are based on
abrupt changes of either the pump power 
\cite{Yi2015, Yi2016, Brasch2016_1, Brasch2016_2}
or the pump frequency \cite{Stone2017arXiv}, 
and referred to as 
``power kicking" or ``frequency kicking",
respectively.
These changes occur
over
timescales that are 
faster than the thermal 
time constant of the micro-resonator.
%
%
Referring to
Fig.~\ref{FIG_setup}(c),
the pump is
typically 
provided 
by a benchtop ECDL
with a piezo-controlled frequency.
The servo feedback 
is applied to the piezo-controller, 
and not to the current source like 
in 
the 
procedure described 
above.
%
%
%
%
For power kicking,
%
the modulator
in Fig.~\ref{FIG_setup}(c)
consists of a
combination of 
acousto-optic and electro-optic modulators
(with modulation frequencies up to $\sim$100 MHz).
%
%
In contrast, the 
frequency kicking protocol
is achieved 
by manipulating the single sideband 
from a quadrature phase shift keying (QPSK) modulator driven 
by a fast-tuning voltage-controlled oscillator (VCO).
%
%
%
In both protocols,
the modulators 
are controlled by an extra SG
and 
have optical insertion loss of a few dB.
An
optical amplifier (OA) 
is thus inevitably required 
to boost
the pump power.
A band-pass filter (BPF) 
is also preferred 
to suppress the undesired ASE
from the OA.
Note that AOMs and QPSKs 
require a specific polarization 
for optimal operation,
so that an extra PC is 
also needed.
These instruments dramatically increase the footprint,
power consumption,
and cost 
of the micro-comb system.
In this work, we are able to eliminate these bulky components
by 
implementing a frequency-kicking protocol
directly with an ECDL.





\section{Results and discussion}

%
%

\newcommand{\widthpage}{170mm}

\renewcommand{\heightimage}{45mm}

\begin{figure*}[h!]
\begin{center}
\includegraphics[width=\widthpage]{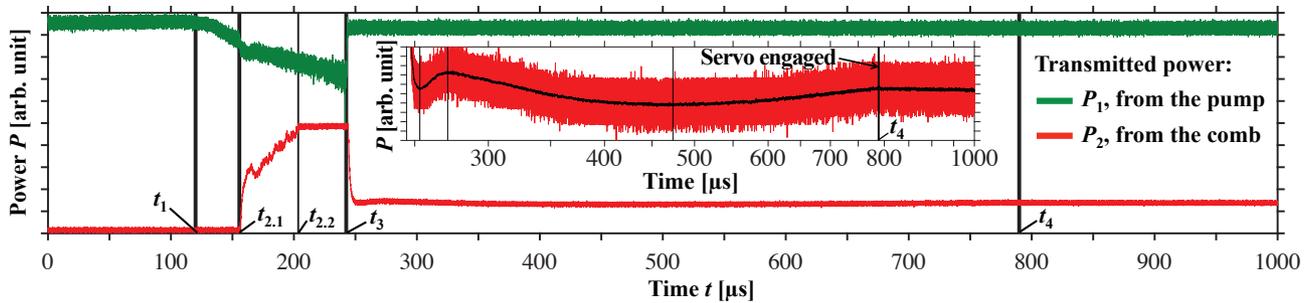}
\caption{
Frequency 
tuning leading to 
the generation of a single soliton,
and its subsequent locking.
The green and the red curves 
are the transmitted power 
from the pump and from the comb, respectively.
%
%
%
%
The inset a close-up view of 
the thermal drift and
the soliton locking.
The black line is a moving average,
and the horizontal scale is logarithmic.
}
\label{FIG_soliton_locking}
\end{center}
\end{figure*}


The ECDL is 
driven
at 350 mA where its output power is 42 mW,
and the power 
coupled into
the tapered fiber 
is 34 mW.
%
%
%
Fig.~\ref{FIG_soliton_locking} 
shows the evolution of
the pump power $P_1$
measured after the micro-resonator,
as the pump frequency $f_{\psub}$
is linearly tuned
from the blue to the red side of 
the resonance.
Also shown is the 
generated
comb power $P_2$.
Up to $t_1$,
$f_{\psub}$
is so far from resonance
that it does not couple significantly to the micro-disk
and 
$P_1$
is relatively constant.
%
This situation changes 
near $t_1$
where
a
reduction of 
$P_1$
is observed
as more light couples to the micro-disk.
%
By
$t_{2.1}$,
the power coupled to the resonator 
eventually reaches 
threshold for parametric oscillation,
leading to 
the onset of a primary comb
and a sharp increase of $P_2$.
%
From $t_{2.1}$ to 
$t_{3}$,
$P_1$
further decreases
as more pump power is 
coupled to the resonator,
%
and between $t_{2.2}$ 
and $t_{3}$,
PD$_2$ is saturated.
Heating of the
resonator induces the overall triangular-shaped profile of
the pump power transmission 
\cite{Carmon2004}.
Upon arriving at the edge of the blue-detuned side, 
the pump frequency
is kicked a few MHz to the red-detuned side for a few \textmu s, 
which induces the soliton state 
in 
the resonator. 
Finally, at $t_{4}$,
\textit{i.e.} a few hundred \textmu s after the scanning, 
the servo loop is engaged to adjust $f_{\psub}$,
so as to
lock the
soliton state indefinitely against thermal drifting.
Variations in 
$P_2$
are further suppressed,
suggesting from Eq.~(\ref{I_sol})
that the 
pump-resonance 
detuning is 
indeed stabilized.


\renewcommand{\heightimage}{45mm}
\begin{figure}[h!]
\begin{center}
\begin{minipage}[b]{\widthminipage}
\centering
\raisebox{49mm}{(a)}
\includegraphics[height=52mm]{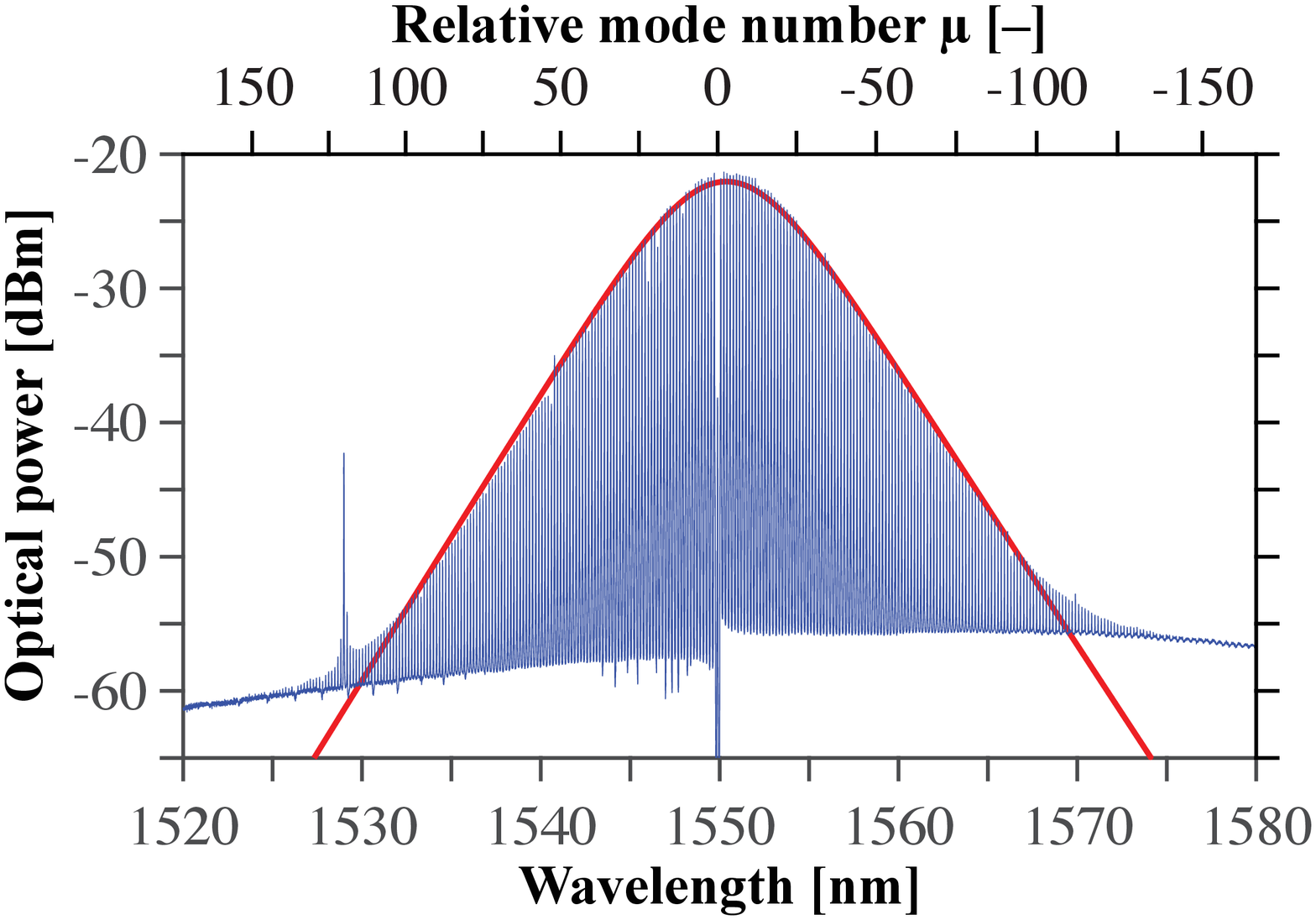} 
\\
\vspace{2mm}
\raisebox{37.5mm}{(b)}
\includegraphics[height=40.5mm]{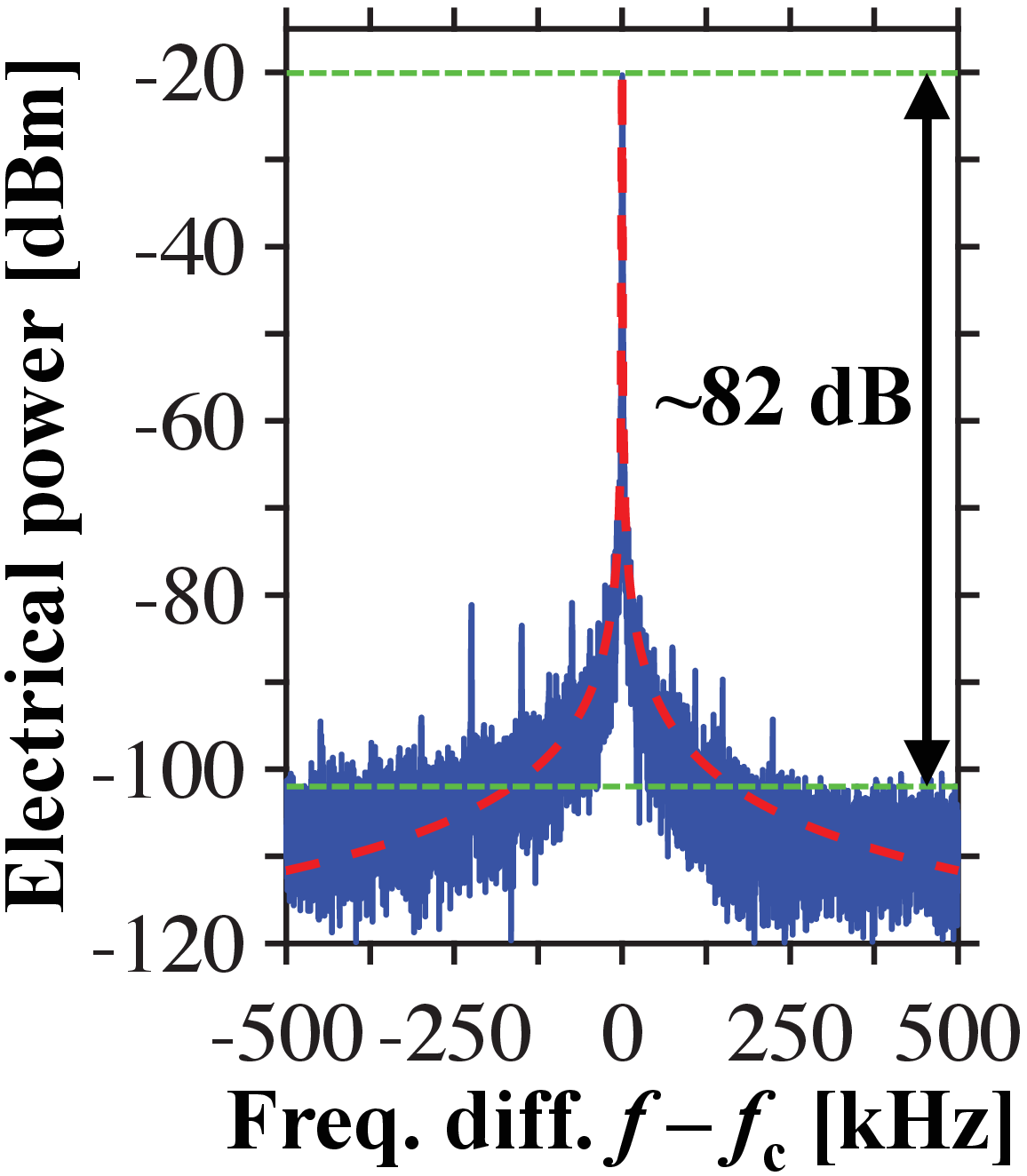}
\raisebox{37.5mm}{(c)}
\includegraphics[height=40.5mm]{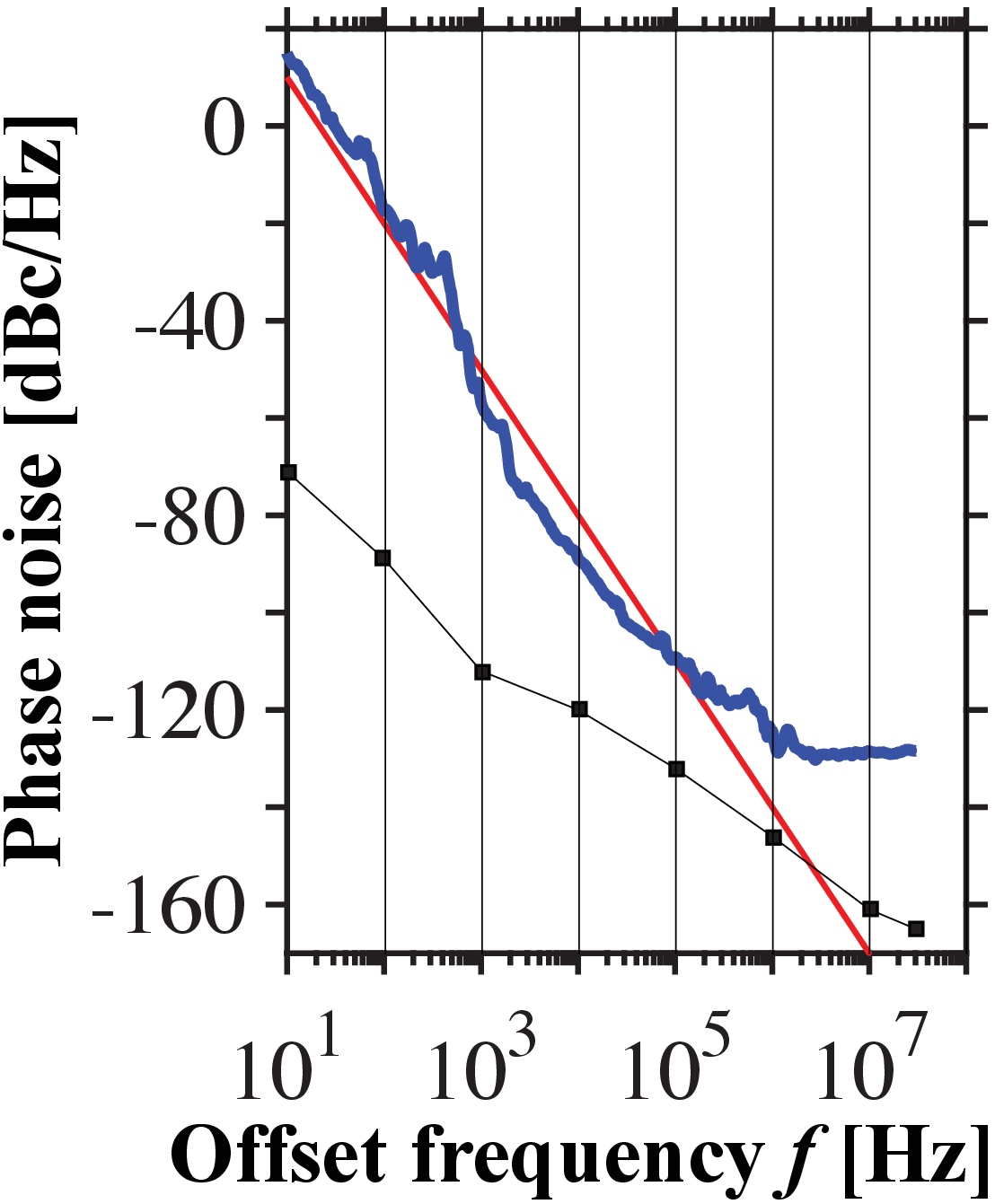}
\caption{
(a) Optical spectrum measured 
in the single-soliton regime.
The red line is a fit according to Eq.~(\ref{F_1}).
%
%
%
%
%
(b) 
Corresponding
RF electrical spectrum,
with $f_{\csub} \approx 22.02$ GHz,
fitted with a Lorentzian (in red).
The bottom green line indicates the noise level.
(c)
Phase noise 
spectral density.
The red line is proportional to $1/f^3$
and black squares indicate the instrument sensitivity.
}
\label{FIG_soliton}
\end{minipage}
\end{center}
\end{figure}

To confirm soliton generation,
the 
light output from the T-port of the FBG 
in Fig.~\ref{FIG_setup}(b)
is sent to an optical spectrum analyser 
(OSA, Yokogawa AQ6370).
It is seen in
Fig.~\ref{FIG_soliton}(a)
that the spectral envelope 
is relatively smooth,
without significant
mode crossings.
However, an intensity spike 
with an amplitude 17 dB above the ASE level
is centered
near 1529 nm
(corresponding to 
a relative mode number $\mu = 120$).
This spectral feature
is a signature of a dispersive wave
\cite{Yi2017},
which can occur from the 
interaction 
of the soliton with 
different transverse modes in the micro-resonator.
%
%
%
%
%
%
%
%
%
%
%
Eq.~(\ref{F_1})
provides an excellent fit for 
the 
remainder
of this spectrum,
and
leads 
$\Delta\tau \approx 196$ fs,
and $\delta \approx 60$ GHz.
%
%
%
The obtained temporal width
is within 
the range of values (125-215 fs) reported in 
\cite{Yi2016_2}.
%
%
%
%
%
%
%
%
%
%
%
%
%
%
%
%
The 
value obtained for $\delta$
indicates a red-shifting 
of the soliton
from the pump
by nearly 3 FSRs.
%
%
This 
self-frequency shift
is attributed to
the spectral recoil 
caused by the dispersive wave,
and possible 
stimulated Raman scattering 
\cite{Yi2016_2, Karpov2016, Yi2017}.
Here,
a single-soliton state is 
directly
generated,
in contrast to \cite{Guo2017}.
No step-like features are visible
in the oscilloscope traces 
of Fig.~\ref{FIG_soliton_locking},
as these would occur
at the transition 
between different soliton states.









The soliton is further confirmed 
by 
assessing
the  
coherence
of the generated comb
\cite{Erkintalo2014}.
Indeed,
a low-noise and narrow RF signal is 
known to be 
a necessary signature of stable soliton formation 
\cite{Herr2014}.
The OSA 
is replaced by a PD 
(50-GHz bandwidth)
connected to 
an electrical spectrum analyser (ESA, 
Rohde \& Schwarz, FSUP26)
%
that 
also measures
the phase noise.
%
%
Fig.~\ref{FIG_soliton}(b)
shows the radio-frequency (RF) electrical spectrum 
recorded 
with a 100-Hz resolution bandwidth.
%
The 
carrier frequency
$f_{\csub} \approx 22$ GHz
corresponds to the 
comb repetition frequency,
\textit{i.e.}
to the beating between neighboring comb lines.
%
These data 
have an SNR
$>$80 dB.
They are fitted with a Lorentzian,
and the 3-dB linewidth is $\sim$25 Hz.
Note that 
the central part is Gaussian
with a 3-dB linewidth of $\sim$1.1 kHz.

%






Finally,
Fig.~\ref{FIG_soliton}(c)
shows the 
phase noise spectral
density 
of the repetition beat note
plotted versus the offset frequency $f$.
%
%
%
%
%
Between 10 Hz and 100 kHz, 
\textit{i.e.} over 4 decades,
the phase noise is 
approximately
proportional to $1/f^3$.
%
%
This
is believed to be contributed by a
combination of pump laser RIN and frequency noise
\cite{Lee2013, Liang2015}.
%
%
%
%
The detected phase noise is smaller than $-100$ dBc/Hz
for offset frequencies higher than 
28 kHz,
which
is comparable to the data
from a previous report
\cite{Yi2017},
where the silica micro-resonators 
were pumped 
with a benchtop ECDL.
%
%
%
%
%
%
%
%
Beyond
1 MHz, 
the detected phase noise reaches 
its 
floor value
of $-129$ dBc/Hz,
attributed to the PD shot noise
\cite{Liang2015}.
%




\section{Conclusion}

Soliton generation was demonstrated in a micro-resonator 
with an 
ultra-low-noise diode laser.
A single-soliton state was successfully stabilized
by locking the frequency of the pump laser 
to the power level of the
soliton comb.
To our knowledge, it is the first time 
a temporal soliton
is realized 
in a 
chip-based 
micro-resonator 
without an optical amplifier.
%
%
This demonstration reduces 
the complexity and cost of soliton experiments,
representing a significant
step towards 
soliton comb systems fully 
integrated on a chip.


\section*{Funding Information}
This research is supported by a DARPA MTO 
DODOS contract (HR0011-15-C-055). 
The views and conclusions contained in this document are those of the authors and should not be interpreted as representing official policies of the Defense Advanced Research Projects Agency or the U.S. Government.
N.V. acknowledges support from the Swiss National Science Foundation (SNSF).

\section*{Acknowledgments}

The authors thank 
Rui-Lin Chao, 
Sarat Chandra Gundavarapu,
and Xinbai Li
for experimental assistance.


\color{black}

\tolerance=1
\emergencystretch=\maxdimen
\hyphenpenalty=10000
\hbadness=10000

\bibliographystyle{lpr}



%
%
%

\end{document}